\documentclass[showpacs,floatfix]{revtex4}
\usepackage{amsmath} \usepackage{amsfonts} \usepackage{amssymb}
\usepackage{graphicx}
\usepackage{epsf}
\usepackage{psfrag}
\usepackage{subfigure}
\usepackage{bm} 
\begin{document}

\title{Dynamics of Two Granules}
\author{Alexandre Rosas, J. Buceta, and Katja Lindenberg}
\affiliation{
Department of Chemistry and Biochemistry,
and Institute for Nonlinear Science
University of California San Diego
La Jolla, CA 92093-0340}
\date{\today}

\begin{abstract}
We study the dynamics of two particles that interact only when in
contact.  In this sense, although not in every particular, the
interactions mimic those in granular materials. The detailed solution
of the dynamics
allows an analysis of the backscattering behavior of the first
particle and of the energy dissipation in the system
as a function of various parameters.   
\end{abstract}
\maketitle

\section{Introduction}
\label{introduction}

Signal propagation in granular materials is a long-standing subject of
interest~\cite{jaeger, sen-book, nesterenko, sinkovits, senpreprint}, and has recently attracted considerable attention in
a number of novel contexts.  One is the observation that 
the attenuation characteristics of signals (impacts) in such materials
might make them good candidates for shock absorption~\cite{sen-pasi}.  Another
is the use of the backscattered signal for the nondestructive
identification of buried objects~\cite{sen98, senpreprint, rogers, naughton, coste}.  

Interactions in granular materials are notoriously complex, but a number
of relatively simple models have been implemented in the context of
signal propagation, models that it is hoped capture the principal
ingredients of the true interactions at least in some regimes.  
A model potential used for a monodisperse chain of particles
that repel upon overlap according to the Hertz
law is given by~\cite{hertz}
\begin{equation}
\begin{array}{l l l}
V(\delta_{i,i+1})&=a\delta^{n}_{i,i+1}, \qquad &\delta\leq 0,\\ \\
V(\delta_{i,i+1})&= 0, \qquad &\delta >0. 
\end{array}
\label{hertz}
\end{equation}
Here 
\begin{equation}
\delta_{i,i+1} \equiv 2R - \left[ (z_{i+1} +x_{i+1}) - (z_i +x_i)
\right], 
\end{equation}
\begin{equation}
a= \frac{2}{5D(Y,\sigma)}\left( \frac{R}{2}\right)^{1/2}, \qquad
D(Y,\sigma) = \frac{3}{2}\left( \frac{(1-\sigma^2)}{Y}\right),
\end{equation}
$Y$ and $\sigma$ denote Young's modulus and Poisson's ratio, $z_i$
denotes the initial equilibrium position of grain $i$ in the chain, and
$x_i$ is the displacement of grain $i$ from this equilibrium position.
The geometric parameter $R$ is the radius if the particles are elastic
spheres.  More generally, $R$ is determined by the principal radii of
curvature of the surfaces at the point of contact~\cite{landau}.
The exponent $n$ is $5/2$ for spheres, it is $2$ for cylinders, and
in general depends on geometry. 

Two features of the potential are decisive for the
resulting propagation properties of the chain: one is the 
exponent $n$, and the other is the absence of an attractive portion in
the potential.  It is in this latter respect that the granular chain
differs profoundly from the many models of nonlinearly interacting
oscillators (such as the Fermi-Pasta-Ulam chain) extensively studied
in recent years~\cite{breatherF,ourchaos}.

Two decades ago Nesterenko~\cite{nesterenko} showed that under appropriate
assumptions, among them the slow spatial variation of the
displacements $x_i$, the
equations of motion for the granular particles could be
approximated by a continuous nonlinear partial differential
equation that admits a soliton solution for
a propagating perturbation in the chain.  More recently it was shown
that strictly speaking these solutions are solitary waves rather than
solitons~\cite{mackay, sen01}. It is also understood that a potential
exponent $n>2$ is required to accommodate solitary waves, i.e.
that a ``hard nonlinearity'' is required~\cite{sen-pasi}.
Nesterenko coined the words ``sonic vacuum'' to stress the
importance of the absence of ``harmonic terms'' ($n=2$) in this analysis
because such terms (and only such terms) are associated with
ordinary sound propagation.
Indeed, as $n\rightarrow 2$ the width of the solitary wave diverges. 
More recently, numerical simulations of \emph{discrete} chains have
provided a great deal of additional information about the important role of
discreteness as well as nonlinearity in the signal propagation
properties.

Thus, while the continuum solitary wave solutions provide enormously helpful
analytic insights, the importance of discreteness in the propagation
properties of such chains should not be underestimated.  Neither should
the fact that even with a ``harmonic'' exponent $n=2$, the chain is
decidedly different from a harmonic system \emph{because there is no
attractive restoring force}.  The resulting nonanalyticity of the potential
leads to profoundly different properties than those observed in an ordinary
harmonic chain.  Indeed, although solitary waves are no longer exact
solutions (and there is no longer a perfect sonic vacuum), there are
extremely long lived quasi-solitary waves that
broaden slowly (whereas a harmonic chain would exhibit rapid
dispersion of energy)~\cite{ourchaos}. 

We offer the above observations in support of the usefulness of
a study of a chain with only repulsive forces even with the choice
$n=2$. A number of 
helpful insights can be achieved from such a study if one is
able to obtain analytic results, and this is at all possible only
with $n=2$. In this paper we carry out such an analysis for the simplest
(and yet extremely interesting) such system, namely, one consisting of
only two masses.  Surprisingly, this simple scenario seems not to have
been analyzed in detail in the literature, although the dynamics of 
pairwise collisions is an important input (and phenomenological assumptions
are made about it) in the study of longer chains~\cite{senpreprint}. 
A two-mass system, being explicitly
integrable, can yield valuable information. In view of some of the
current applications of energy propagation in granular materials,
we are particularly interested in \emph{backscattering}
and \emph{energy dissipation} in
the presence of friction, and on the effects of frictional disorder on
these quantities.  The results of our analysis promise to
provide helpful phenomenological input into the propagation of
excitation pulses in longer chains as long as the pulses remain narrow.
For example,
recent experiments on impulse dispersion in tapered chains indicate the
significant role of friction in the propagation and dispersion
process~\cite{nakagawa}.
We find that a two-mass system leads to some unanticipated
nonmonotonic (albeit, in hindsight understandable) behaviors as
one varies system parameters. To support the relevance of our results to
the spherical granular problem, we also present some numerical results for the
Hertz potential with $n=5/2$ and show the similarity in the behavior of
this system and that of the $n=2$ case.

We consider two finite-sized particles (``granules'') of equal mass
$m$. The specific value of the mass does not matter since it
can be scaled out by redefining variables; the only important point is
that the masses are equal.
When the granules ``do not touch'', each moves independently of the other
(i.e., their potential of interaction is zero beyond a certain
distance between their centers).  When they {\em do} ``push
one into the other'', there is a repulsive linear force between
their centers with force constant $k$; the force vanishes once the
particles separate again. 
One or both of the granules are subject to friction, which may be
kinetic (i.e. constant, see Sec.~\ref{model1}) or
hydrodynamic (i.e. proportional to velocity, see Sec.~\ref{model2}).
In order to initiate energy transfer between the granules, one can, for
example, introduce some precompression in the system, or one
can impart one of the granules an initial velocity toward the other.
We do the latter: initially the granules are assumed to be just
touching, the first granule has initial velocity $v_0$ toward the
second, and the second is at rest.  We are interested in 
the velocity of the two granules and the total energy in the system
\emph{immediately after the collision}.  
Note that in the absence of friction
the collision is elastic so that the energy and momentum are conserved.
In this case the first granule simply ends up at rest and the second
with velocity $v_0$, i.e., there is no backscattering.  Of course there
is backscattering, even in the absence of friction, in a longer
chain.

Even these simple systems present a broad range of outcomes, some
more interesting than others, and not all of which will be covered in
detail.  For example, one or the other or both granules may stop moving
altogether before the collision is over.  However, we are particularly
interested in situations where the second granule is still moving
forward after the collision (propagation) and the first granule is
moving backward (backscattering). Our analysis will focus on these
outcomes and the conditions that lead to them.

In Sec.~\ref{model1} we present and solve the two-particle model with
kinetic friction.  Section~\ref{model2} presents a
similar calculation for the model with hydrodynamic friction.
A summary of results, and a brief outlook of future
work, are gathered in Sec.~\ref{summary}.

\section{Two Particles with Kinetic Friction}
\label{model1}

In this section we examine granule collision when one or both of
the granules are subject to
kinetic friction with friction coefficient $\mu$.
Kinetic friction arises from contact forces when a solid body moves along
a solid surface, and depends on the velocity only in that its direction
is always opposed to the velocity.
Two cases are presented analytically (at least in part). They
turn out to exhibit rather different
behaviors, which we anticipate here as a point of reference.  In
one, only the second granule
is subject to friction.  We find that for a certain range of friction
coefficient values there is both propagation and backscattering, that 
the velocities of the forward moving second granule and the backward
moving first granule are monotonic functions of the friction, but that
the total energy dissipation during the collision is nonmonotonic.
In the second model only the first granule is subject to friction.
Here we again find a range of values of the friction coefficient for
which there is both propagation and backscattering.  Now, however, the
granule velocities are nonmonotonic functions of the friction while the
energy loss is a monotonic function.  Results for the more general case
of both granules subject to (possibly different) friction are
presented numerically.  In this way we cover the
entire range of ``frictional disorder'', as might occur if
the granules are made of different materials.   While most of our discussion
revolves around the case $n=2$, some results for $n=5/2$
are presented.  Also, as a backdrop we keep in mind the most
extreme case of a hard interaction, a hard sphere
($n\rightarrow\infty$).  The interaction  is instantanous in
this case and leads to elastic collisions in which energy and momentum
are conserved. The first granule stops and the second granule moves forward.
This comment emphasizes the observation that only collisions of finite
duration can lead to a combination of backscattering and propagation in
two granules.

We rescale variables
(see Appendix~\ref{appendixa}) so
that we can set the coefficient $a\equiv 1$ in Eq.~(\ref{hertz})
regardless of the value of $n$.  This rescaling
allows presentation of results in terms of a single parameter, the
scaled friction.

\subsection{Frictionless First Granule}
\label{firstgranule}

Consider first the case of a frictionless first granule.  The 
movement of the second granule will then only start if the elastic
compression exceeds the frictional force, $x_1^{n-1}>\eta$,
and so at first
only the first granule moves and its motion is governed by the equation
of motion
\begin{equation}
\ddot{x}_1(t) = - [x_1(t)]^{n-1}.
\end{equation}
The initial condition for this problem is
$x_1(0)=0$ and $\dot{x}_1(0) = 1$. The solution for $n=2$ is
easily found to be
\begin{equation}
\label{eq:sol1}
x_1(t) = \sin(t),
\end{equation}
and is valid until the time $t_0$ at which $x_1(t_0)=\eta$, 
that is, until $t_0 = \arcsin(\eta)$.

At time $t_0$ the second granule starts moving as well, and the system is
now governed by the two coupled equations of motion
\begin{subequations}
\begin{eqnarray}
\ddot{x}_1({t'}) & = & - \left [ x_1({t'}) - x_2({t'})\right ] 
\\ \nonumber\\
\ddot{x}_2({t'}) & = & - \eta  - \left [ x_2({t'}) - x_1({t'})
\right ]
\end{eqnarray}
\end{subequations}
with the initial conditions: $x_1(t'=0)= \eta$, $\dot{x}_1(t'=0) =
\sqrt{1 - \eta^2}$, and $x_2(t'=0)=\dot{x}_2(t'=0)=0$. Here
we have defined ${t'}=t-t_0$.
The solution is found to be:
\begin{subequations}
\begin{eqnarray}
x_1({t'}) & = &
 \frac{1}{4}
     \left[ - \eta (-3 + {t'}^2)
         + 2 (1-\eta^2)^{1/2}t'   + \eta \cos ({\sqrt{2}}t') +
        {\sqrt{2}} (1-\eta^2)^{1/2}\sin ({\sqrt{2}} t') \right ]
\label{eq:x1} \\
\nonumber\\
x_2({t'}) & = &
- \frac{1}{4 }
    \left[ \eta (-1 + {t'}^2 ) - 2 (1-\eta^2)^{1/2}{t'}
     +    \eta \cos ({\sqrt{2}} t') +
      {\sqrt{2}} (1-\eta^2)^{1/2}\sin ({\sqrt{2}} t') \right ] .
\label{eq:x2}
\end{eqnarray}
\end{subequations}

\begin{figure}
\begin{center}
\resizebox{7cm}{!}{\includegraphics{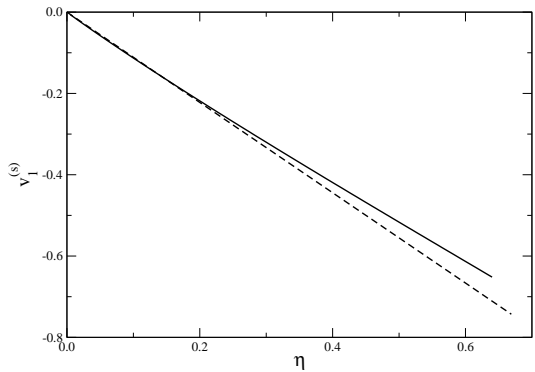}}
\caption{Velocity of the first (frictionless) granule at the end of the
collision as a function of friction.
Solid curve: $n=2.5$
(Hertz potential for spheres). Dashed curve: $n=2$. \label{fig:velocity}}
\end{center}
\end{figure}

If $\eta$ is too large, the second granule will stop before the two
granules lose contact.  For sufficiently small $\eta$ this
does not happen and the above
solutions then remain valid until contact is lost.
The time $t'=t_1$ when this occurs is the time at which
$\Delta x (t_1)= x_1(t_1) - x_2(t_1)$ vanishes, that is, when
$t_1 = \pi/\sqrt{2}$. 
The velocities of the granules at the moment of separation $t=t_0+t_1$ are
\begin{subequations}
\begin{eqnarray}
v_1^{(s)} &=& -\frac{\eta \pi}{2\sqrt{2}} \\
v_2^{(s)} &=& \left ( (1-\eta^2)^{1/2} -\frac{\eta \pi}{2\sqrt{2}} \right ).
\end{eqnarray}
\end{subequations}
The values of $\eta$ for which this description is valid
are those for which $v_2^{(s)}>0$,
that is, for $0<\eta<(1+\pi^2/8)^{-1/2} = 0.669\ldots$ 
Beyond time $t=t_0+t_1$ each granule continues on its course, the
first forever and the second until friction brings it to rest.  
Backscattering and propagation thus occur when $\eta<0.669\ldots$
As a function of $\eta$ in this range, the magnitude of the backward
velocity increases monotonically while that of the forward velocity 
decreases monotonically. The backscattering is shown in
Fig.~\ref{fig:velocity}, which we exhibit in order to stress the similarity
in the behavior of the $n=2$ potential and that of the
Hertz potential for spheres.
As $n$ increases the range of $\eta$ for which both $v_1^{(s)}$ and
$v_2^{(s)}$ are nonzero shrinks, until it shrinks away completely for a
hard sphere potential.

Perhaps the most dramatic outcome of this calculation is the fact that
\emph{the friction on the second granule causes the first granule
to move backward} (recall that if both granules are frictionless the
first one simply stops). In other words, the friction on the second
granule is responsible for backscattering. This behavior, strange at
first glance, occurs because the first granule ``stops too
soon'', that is, it stops before the spring has had time to decompress
completely.  Indeed, if the friction on the second granule is so strong
that it does not move at all, the first will move backward with velocity
$v_1=-1$. Our model lies between the limiting cases of zero
and very large $\eta$ ($-1 \le v_1^{(s)} \le 0$).  

\begin{figure}
\vspace*{0.2cm}
\begin{center}
\resizebox{7cm}{!}{\includegraphics{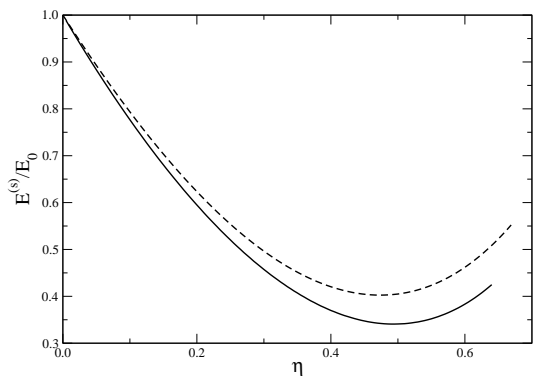}}
\caption{Ratio of the total energy at the moment of separation to
the initial energy as a function of friction. Solid curve: $n=2.5$
(Hertz potential for spheres). Dashed curve: $n=2$. \label{fig:energy}}
\end{center}
\end{figure}

In Fig.~\ref{fig:energy} we show the ratio of the total kinetic energy
of the system at the end of the collision to the initial energy,
\begin{equation}
\frac{E^{(s)}}{E_0} = 1 -\frac{\eta \left ( 1 - \eta^2 \right )^{1/2}
\pi}{\sqrt{2}}+ \left (\frac{\pi^2}{4}-1 \right )\eta^2.
\end{equation}
In the rescaled variables, $E_0=1/2$.
Curiously, there is a region where this ratio \emph{increases}
with increasing friction coefficient. 
The reason for this behavior is that when friction increases beyond
a certain point (beyond the minimum in the curve), relatively more
energy is transfered to the first granule (as opposed to being simply
dissipated) by the end of the collision.  Again we also show the numerical
results obtained for the Hertz potential with $n=5/2$
(solid curve) and comment on
the similarity with the $n=2$ curve.  

\subsection{Frictionless Second Granule}
\label{secondgranule}

Next we consider the case of a frictionless second granule.  The first
granule now moves forward, eventually stops, and \emph{perhaps}
moves backward, while the second granule moves forward. 

As before, the granules are initially just touching, the first granule
having an initial velocity $v_0\equiv 1$.  While the
first granule is moving forward the system is governed by the equations 
\begin{subequations}
\begin{eqnarray}
\ddot{x}_1({t}) & = & -\eta - \left [ x_1({t}) -
x_2({t})\right ]\label{first}\\ \nonumber\\
\ddot{x}_2({t}) & = & -\left [ x_2({t}) - x_1({t}) \right ].
\label{second}
\end{eqnarray}
\end{subequations}
Explicit integration yields
\begin{subequations}
\begin{eqnarray}
x_1(t) & = &
\frac{1}{4}
\left[-\eta + 2 t - \eta t^2 +
      \eta \cos ({\sqrt{2}} t) +
      {\sqrt{2}}\sin ({\sqrt{2}} t)\right]
\label{eq:x1model2}\\ \nonumber\\
x_2(t) & = &
\frac{1}{4}
\left[\eta + 2 t -
        \eta t^2 -
        \eta \cos ({\sqrt{2}} t) -
        {\sqrt{2}} \sin ({\sqrt{2}} t)\right].
\label{eq:x2model2}
\end{eqnarray}
\end{subequations}
These equations hold as long as the velocity $v_1(t)=\dot{x}_1(t)>0$.
This velocity is given by
\begin{equation}
v_1(t) =  \frac{1}{4}\left[ 2 - 2\eta t +
        2\cos ({\sqrt{2}} t) -
        {\sqrt{2}}\eta\sin ({\sqrt{2}} t)
       \right] .
\end{equation}
The time $t_0$ at which $v_1(t_0)=0$ cannot be found explicitly in
general, but the problem may still be treated 
analytically for small or for large $\eta$, or numerically for
arbitrary $\eta$.

\begin{figure}
\begin{center}
\resizebox{7cm}{!}{\includegraphics{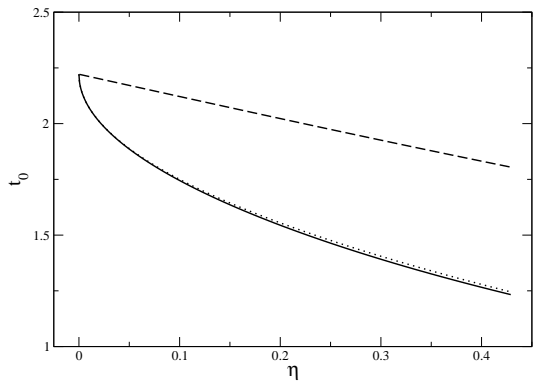}}
\caption{ Comparison of the small-$\eta$ analytical solution (dotted) and 
the numerical solution (solid) of the stopping condition $v_1(t_0)=0$. 
The dashed line is the time it would
take the granules to separate if the solutions~(\ref{eq:x1model2}) and
(\ref{eq:x2model2}) remained valid until then.  The fact that it lies
above the other curves indicates that the first granule stops before the
collision is over.  \label{fig:timecompare2}}
\end{center}
\end{figure}

If $\eta \ll 1$, we can expand $v_1(t)$ around $t=\pi/\sqrt{2}$
(the solution for $\eta =0$). We find that
\begin{equation}
t_0 =  \frac{\pi }{\sqrt{2}} - \left(\frac{\pi \eta}
{\sqrt{2}}\right)^{1/2} + {\cal{O}}({\eta}^{3/2}). \label{eq:t0model2}
\end{equation}
In Fig.~\ref{fig:timecompare2} we compare this analytical approximation
with the numerical solution. The small-$\eta$ 
approximation is clearly very good for a rather large
range of $\eta$.  Therefore, we proceed using this approximation.
In the figure we also show the time that it would take the collision to
be over if the solutions~(\ref{eq:x1model2}) and (\ref{eq:x2model2})
would remain valid, i.e., the solution $t_0'$ of the equation
$x_1(t_0')=x_2(t_0')$.  This time, shown as a dashed line, is always above
the time at which the first granule stops, indicating that the first
granule {\em necessarily} stops before the collision is over. This persists
even for large $\eta$: as
$\eta\rightarrow\infty$ the stopping time of the first granule goes to
zero as $1/\eta$, while the time for the collision to be over if the
above solutions would remain valid goes to zero as $2/\eta$. 

When the first granule stops, the equations of motion are no longer 
Eqs.~(\ref{first}) and (\ref{second}).  The initial conditions for the
new set of equations are the velocities and positions found above at
time $t=t_0$.  Within the small-$\eta$ approximation we use
\begin{subequations}
\begin{eqnarray}
x_1(t_0)&=&\frac{1}{8}\left[ 2\sqrt{2}\pi -(\pi^2+4)\eta
\right] \label{ic1}\\ \nonumber\\
x_2(t_0)&=&\frac{1}{8}\left[ 2\sqrt{2}\pi -
4(2^{3/4})\sqrt{\pi \eta} -(\pi^2-4)\eta \right]
\label{ic2}\\ \nonumber\\
v_1(t_0)&=&0 \label{ic3}\\ \nonumber\\
v_2(t_0)&=& \frac{1}{2}\left[ 2-\sqrt{2}\pi\eta
\right] \label{ic4}
\end{eqnarray}
\end{subequations}
as the initial conditions.

The equations of motion at this point depend on whether the first granule
simply remains stationary or starts moving backward.  The latter occurs
if the elastic force is greater than the friction, that is, if
\begin{equation}
[x_1(t_0)-x_2(t_0)] =  
\frac{1}{2}
 \left[-\eta - \eta\,\cos (2^{1/4}\sqrt{\eta\pi}) + {\sqrt{2}}\,
     \sin (2^{1/4}\sqrt{\eta\pi})\right] >  \eta.
\end{equation}
Therefore, for $\eta$
smaller than a critical value ($\eta=0.439\ldots$) the first granule
will start moving
backward, and we confine our analysis to this case. 
The equations of motion then are (note that since the first
granule is moving backward, the friction is also reversed):
\begin{subequations}
\begin{eqnarray}
\ddot{x}_1({t'}) & = & \eta -  \left [ x_1({t'}) -
x_2({t'})\right ]
\label{eq:motionx1}\\ \nonumber\\
\ddot{x}_2({t'}) & = & - \left [ x_2({t'}) - x_1({t'}) \right ]
\label{eq:motionx2}
\end{eqnarray}
\end{subequations}
where $t' = t - t_0$. The initial conditions are those given in
Eqs.~(\ref{ic1})-(\ref{ic4}).

\begin{figure}
\begin{center}
\resizebox{7cm}{!}{\includegraphics{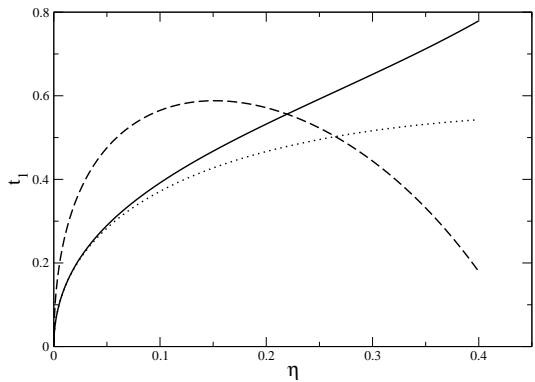}}
\caption{Comparison of time required for the two granules to
separate calculated
numerically (solid) and analytically for small $\eta$ (dotted). The time
required for the first granule to stop according to the equations of
motion~(\ref{eq:motionx1}) and (\ref{eq:motionx2})
is shown as a dashed curve. Therefore, if $\eta
\lesssim 0.22$ the first granule is still moving backwards at the end of
the collision.
\label{fig:timecompare3}}
\end{center}
\end{figure}

The solutions of this set as a series in $\eta$ can easily be exhibited
explicitly but are not very instructive. Two cases must be
distinguished: 1) The first granule is still moving backward when the
collision ends; and 2) The first granule stops before the collision
ends, at which point the equation of motion~(\ref{eq:motionx1}) no longer
holds.  In either case, from the series solutions one can
calculate the time $t'=t_1$ at which the collision ends as the solution
of the equation $x_1(t')=x_2(t')$.  If we assume that the first granule
is still moving we find
\begin{equation}
t_1 =  \left( \frac{\pi \eta}{\sqrt{2}}\right)^{1/2}
-\eta+{\cal{O}}(\eta^{3/2}).
\label{omega1}
\end{equation}
This solution is shown as the dotted curve in Fig.~\ref{fig:timecompare3},
as is the numerically obtained exact
solution (solid line).  The approximation (\ref{omega1}) begins
to seriously deviate from the exact solution at around $\eta \sim 0.1$. 
The dashed curve indicates the time at which the
first granule stops before the end of the collision
if Eq.~(\ref{eq:motionx1}) remains valid until it
does.  This analysis is therefore appropriate and leads to
backscattering (and propagation) if $\eta\lesssim 0.22$.

\begin{figure}
\vspace*{0.2cm}
\begin{center}
\resizebox{7cm}{!}{\includegraphics{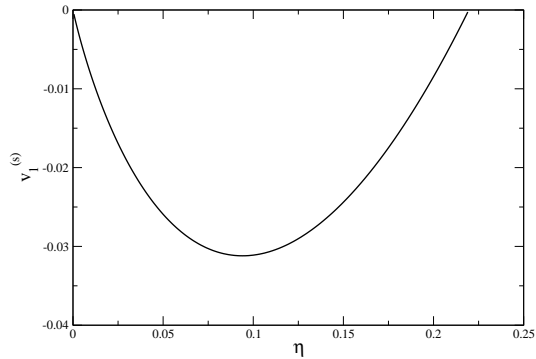}}
\caption{ Final velocity of the first granule.\label{fig:finalv1}}
\end{center}
\end{figure}

In Fig.~\ref{fig:finalv1} we see that as $\eta$ increases the final
backward velocity of the
first granule is nonmonotonic, that it, is increases from zero to
approximately $3\%$ of the
initial velocity, and then decreases again.
In this model, we thus observe that there is backscattering for small $\eta$.
For $\eta =0$ and for $\eta$ greater than a certain value ($\gtrsim 0.22$),
the first granule is at rest when the collision ends and there is no
backscattering.
Therefore, the backscattering in this model happens because the
\emph{first} granule (as opposed to the second granule in the previous
example) ``stops too soon'' (i.e. before the collision is over),
but not ``too late'' (i.e. when the compression is still strong enough
to overcome friction).  The required balance leads to the nonmonotonic
behavior seen in Fig.~\ref{fig:finalv1}.
Even though the backscattering is small (e.g. compared to that in the
first model), it is conceptually striking.  However, the total energy
of the system at the end of the collision is a monotonically decreasing
function of $\eta$, in contrast to the behavior in the previous case
shown in Fig.~\ref{fig:energy}.

\subsection{Granules with Arbitrary Friction}
\label{arbitrary}
While the results presented above deal with the extremes of 
frictional disorder, it is helpful to present a broad brush of graphical
results
for arbitrary combinations of $\eta_1$ and $\eta_2$, the friction
parameters for the two granules. This is also an opportunity for
re-assertion that these broad brush results are essentially the same for
the $n=2$ and the $n=5/2$ Hertz potentials for two granules.
Since the graphical results are necessarily numerical, we present them for
the $n=5/2$ potential. They are qualitatively indistinguishable from
those for $n=2$.

\begin{figure}[htb]
\begin{center}
\epsfxsize = 2.in
\epsffile{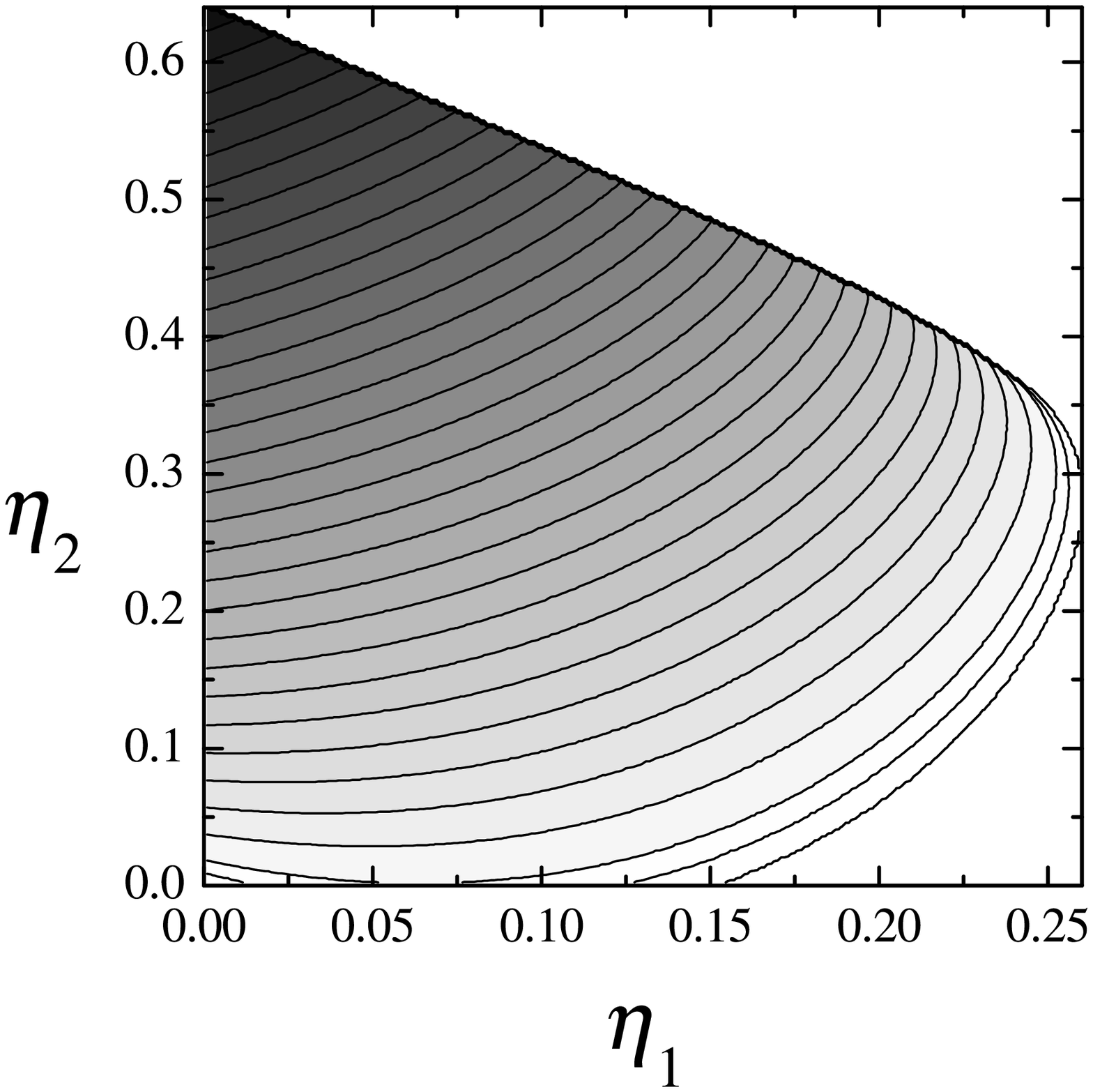}
\epsfxsize = 2.in
\epsffile{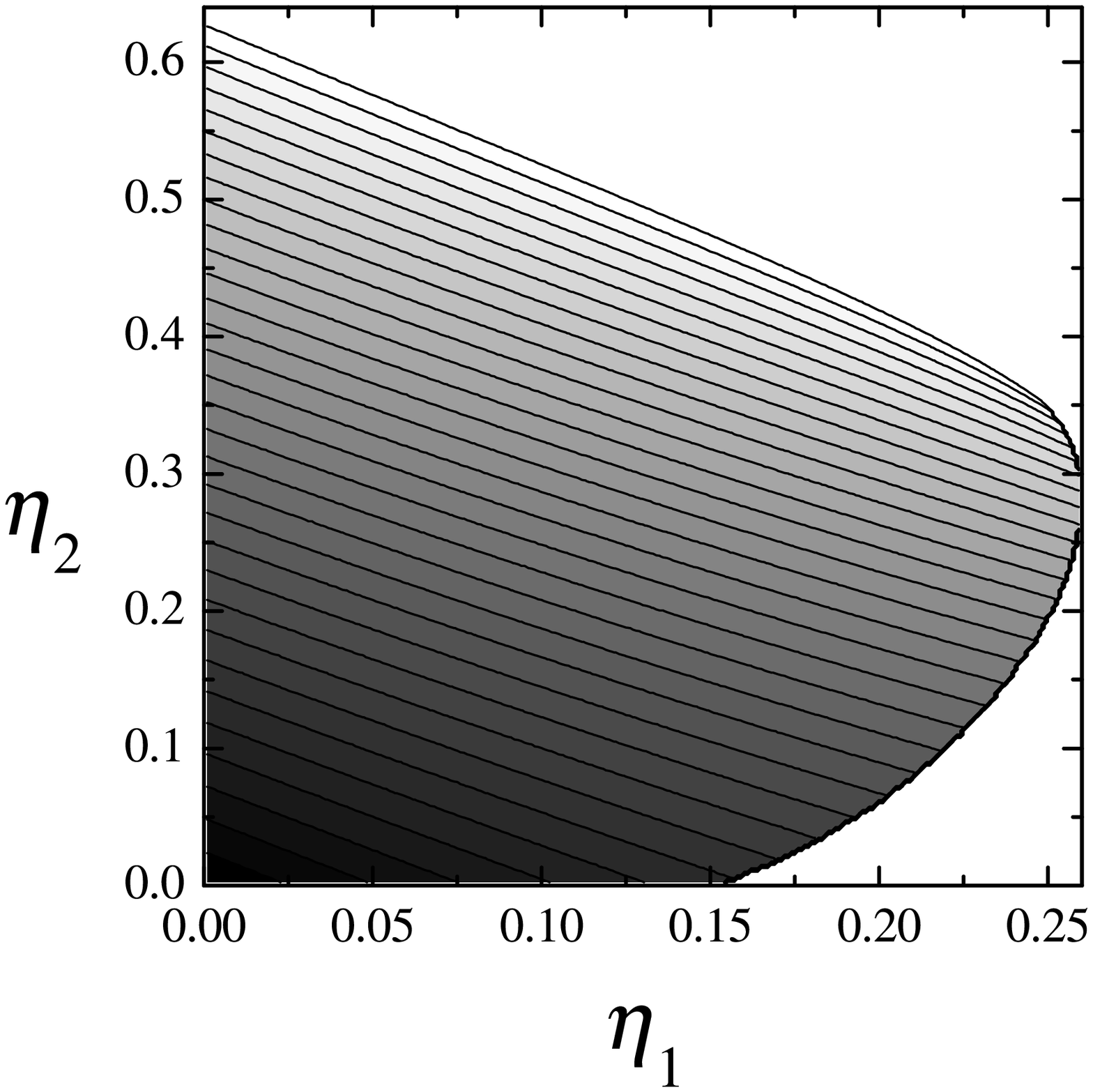}
\epsfxsize = 2.in
\epsffile{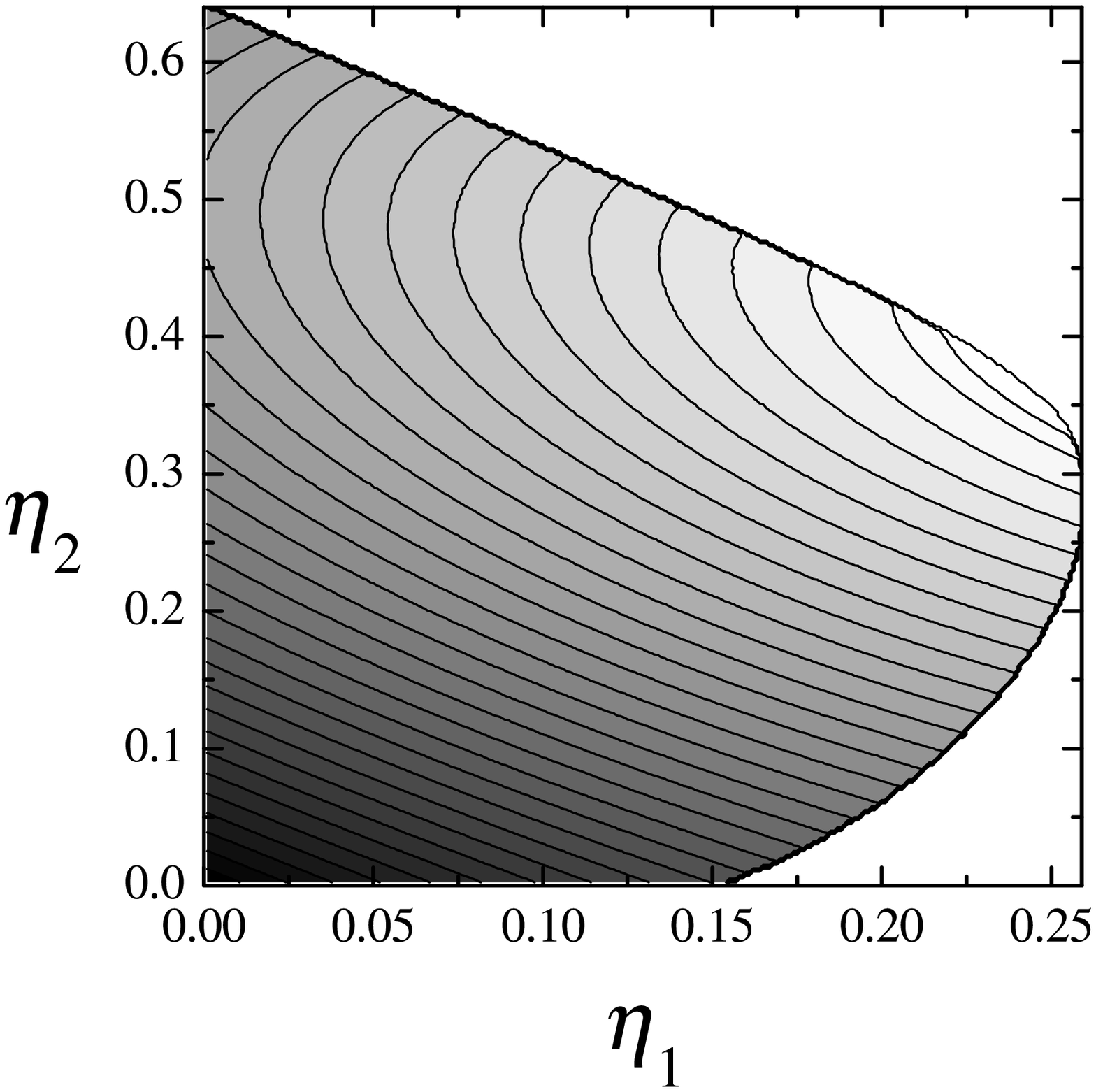}
\end{center}
\caption{
Backscattering velocity $-v_1^{(s)}$ (first panel), forward velocity
$v_2^{(s)}$ (second panel), and energy ratio $E^{(s)}/E_0$ (third panel)
at the end of a collision for the Hertz potential for spheres. 
In the first two 
the gray scale shows darker coloring for higher speeds.  In the third
panel darker coloring indicates higher energy.  The curves in the first
two panels are iso-velocity lines, and in the third panel iso-energy
lines.}
\label{fig:hertz}
\end{figure}

In Fig.~\ref{fig:hertz} we show in gray scale the magnitudes of
$v_1^{(s)}$ (first panel), $v_2^{(s)}$ (second panel) and energy ratio
$E^{(s)}/E_0$ at the end of the collision
in the $(\eta_1,\eta_2)$ regime where there is simultaneous
backscattering and propagation. Darker shading represents a higher speed
(first and second panels) and a higher energy (third panel).  
Figure~\ref{fig:velocity}
is associated with an upward trajectory in the first panel at
$\eta_1=0$.  The monotonically increasing backscattering seen in
Fig.~\ref{fig:velocity} corresponds to the upward darkening.
Figure~\ref{fig:energy} is associated with the same upward trajectory, now
in the rightmost panel.
The nonmonotonicity is seen in the lightening followed by darkening.  
The nonmonotonicity of Fig.~\ref{fig:finalv1} is associated with the
shading in the
first panel associated with the horizontal trajectory along $\eta_2=0$,
captured in the curvature of the iso-velocity lines.
The monotonicty of $v_2^{(s)}$
discussed in the $(\eta_1,0)$ and $(0,\eta_2)$ cases is evident in the
middle panel. 

This rather complex behavior
arises from the interplay of energy dissipation and energy transfer
during a collision.  Thus, for example, increasing friction on the first
granule causes it to stop sooner.  If it stops too soon (``high
$\eta_1$''), the spring is
not sufficiently compressed to overcome friction and it may happen that
the whole process simply stops then and either one or the other or both
granules do not move.  On the
other hand, if it stops too late (``low $\eta_1$'') then the energy transfer
to the second granule may cause it to move forward, but there may not be
sufficient energy to cause backscattering. Or backscattering may occur
but not forward motion. Between these situations (``intermediate
$\eta_1$''), both
granules may be in motion at the end of the collision.  

\section{Two Particles with Hydrodynamic Friction}
\label{model2}

When dissipation is of hydrodynamic origin, the
dissipative force is proportional to the particle velocity, $-\gamma
\dot{x}$.  Hydrodynamic friction arises from the motion of a body
against a lubricated surface or in a liquid or gaseous medium.
We find that the cases of equal and slightly unequal
friction coefficients can be solved analytically and cover essentially
the whole spectrum of possible behaviors.

The scaled equations of motion (see Appendix~\ref{appendixa})
for the potential with $n=2$ are
\begin{subequations}
\label{eq:model5}
\begin{eqnarray}
\ddot{x}_1({t}) & = & -\gamma_1 \dot{x}_1 -  \left [ x_1({t})
- x_2({t})\right ] \\ \nonumber\\
\ddot{x}_2({t}) & = & -\gamma_2 \dot{x}_2 - 
\left [ x_2({t}) - x_1({t}) \right ]
\end{eqnarray}
\end{subequations}
Even though these
equations can be integrated exactly, the time required for the two
granules to
separate can, in general, only be calculated numerically. Here we consider
the case $\gamma_1 \approx \gamma_2$, for which we can find an
approximate analytic expression for this time. Thus, we set
$\gamma_i = \gamma(1 + \delta_i)$ with $\delta_1=-\delta_2\equiv \delta$
and solve the problem for small $|\delta|$.

Defining $x_\pm = x_1 \pm x_2$, we can 
obtain from Eqs.~(\ref{eq:model5}) the associated Laplace transforms
$X_\pm(s)$:
\begin{subequations}
\label{eq:model5lt}
\begin{eqnarray}
 \delta\gamma  s X_-(s)+  \left(\gamma  s + s^2
\right)X_+(s)  =1 \\ \nonumber\\
  \left ( 2  + \gamma s + s^2 \right ) {X_-(s)} +
\delta \gamma  s X_+(s) = 1.
\end{eqnarray}
\end{subequations}
The solution of these equations to first order in $\delta$ is
\begin{subequations}
\begin{eqnarray}
X_+(s)&=&
\frac{1}{s(\gamma + s)} - \frac{\delta \gamma}{s
     {\left( \gamma + s \right) }
     \left[ 2  + s \left( \gamma + s \right)
       \right] } \\ \nonumber\\
X_-(s)&=&
   \frac{1}{2  + s \left( \gamma + s \right) }-
   \frac{\delta \gamma}{\left( \gamma + s \right)
     {\left[ 2  +
         s \left( \gamma + s \right)  \right] }}.
\end{eqnarray}
\end{subequations}
The time functions $x_\pm (t)$ are the inverse Laplace transforms of
$X_\pm(s)$. Furthermore, the time $t_0$
at which the granules separate can be
calculated by solving the equation $x_-(t=t_0)= 0$. 
To zeroth order in $\delta$ (that is, for $\delta=0$), we have
\begin{equation}
x_-(t_0) =   e^{-\gamma t_0/2}  \frac{\sin (\omega t)}{\omega} =0,
\end{equation}
where $\omega = \sqrt{8  - \gamma^2}/2$. If $\gamma^2\geq 8$ the
collision never ends.  If $\gamma^2<8$ then it ends at
time $t_0 = \pi/\omega$.  To calculate the separation time to first
order in $\delta$,
we write $t_1 = t_0 + \epsilon/\omega$, expand $x_-(t_1)=0$ to first
order in $\epsilon$, and solve the resulting equation for $\epsilon$, to
obtain
\begin{equation}
\epsilon = -\frac{\delta\gamma\omega}{2} \left(
1+e^{-\gamma\pi/2\omega}\right).
\end{equation}

\begin{figure}
\begin{center}
\resizebox{8cm}{!}{\includegraphics{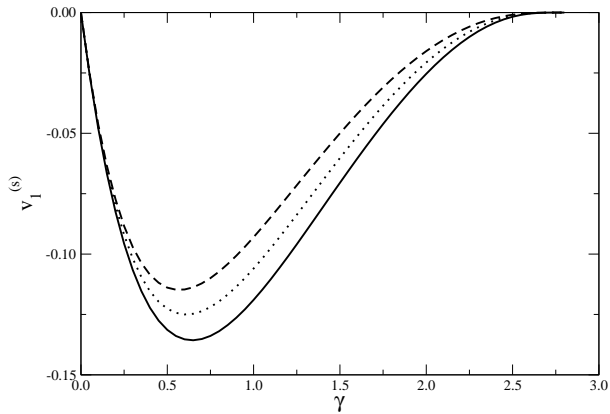}}
\caption{Final velocity of the first granule when $\gamma_1>\gamma_2$,
$\delta = 0.1$ (dashed), $\gamma_1=\gamma_2$, $\delta = 0$ (dotted), and
$\gamma_1<\gamma_2$, $\delta = -0.1$ (solid).
\label{fig:finalv1m5}} \end{center}
\end{figure}

The final velocities of the granules at the end of the collision to
first order in $\delta$ are:
\begin{subequations}
\begin{eqnarray}
v_1^{(s)} & = &  \frac{1}{2}\left(1 - e^{\gamma\pi/2 \omega } \right) 
      e^{-\gamma\pi/\omega}
+ \frac{\delta\gamma^2}{4}
\left(1 + 2 e^{\gamma \pi/ 2\omega }+ e^{\gamma \pi /\omega } 
\right)  e^{-3 \gamma \pi/2 \omega } \\
v_2^{(s)} & = &  \frac{1}{2}\left(1 + e^{\gamma\pi/2 \omega } \right) 
      e^{-\gamma\pi/\omega}
+ \frac{\delta\gamma^2}{32\omega^2}
\left[8\omega^2 +
   \left(12\omega^2 -\gamma^2+ 8 \right) e^{\gamma \pi/2 \omega }
+             \left(4\omega^2 -\gamma^2  + 8 \right)
e^{\gamma \pi/\omega }\right] e^{-3 \gamma \pi/2 \omega} .
\end{eqnarray}
\end{subequations}
We immediately see that the backscattering is
greater if $\gamma_2>\gamma_1$ ($\delta>0$) than if $\gamma_1$ is the
larger friction coefficient. This behavior is similar
to that which occurs when we have kinetic friction.
Thus, again a force in the second granule is more
effective in producing the backscattering (as expected).
In Fig.~\ref{fig:finalv1m5}, we present the velocity of the first granule
at the end of the collision as a function of $\gamma$
for different values of $\delta$.
Once again we find (for each value of $\delta$) an optimal scaled
dissipation constant for which the backscattering is greatest. 
We note that the total energy at the end of the collision is a
monotonically decreasing function of the average dissipation parameter
$\gamma$.

\section{Summary and Outlook}
\label{summary}

We have presented an overview of the dynamics of two granules subject
to repulsive but not restoring forces (as in granular materials), and to
kinetic or hydrodynamic friction.  In particular, we explored parameter
regimes that would give rise to backscattering \emph{and}
propagation at the end of a collision, with a view toward future studies
of chains of granules subject to friction.  
We explored the consequences of ``frictional
disorder'' in the two-granule system and found dramatically
asymmetric behavior, that
is, backscattering and propagation are affected differently by the
friction on the first and second granules.  For example, 
while friction that is ``too high'' on either granule will lead to
situations where backscattering or propagation or both do not occur, the
range of friction parameters that \emph{do} lead to the occurrence of
backscattering \emph{and} propagation is considerably greater for the second
granule than for the first.  While this is perhaps not
surprising, the nonmonotonicities in the final velocity and energy as
a function of friction point to the interesting dynamical asymmetries in
this problem. 
While most of our results have been calculated for a potential with
$n=2$ [cf. Eq.~(\ref{hertz})], we have argued and shown numerically that
the two-granule results are qualitatively similar for other values
of $n$ (albeit not quantitatively equal). They therefore 
provide some insights into the dynamics involving more realistic
potentials.

Finally, we stress that a chain of many granules differs in important
ways from the two-granule system.  For example, with $n>2$ a
frictionless chain may
support solitons while the $n=2$ chain does not~\cite{nesterenko}. 
Also, due to the
interactions with further granules, there may be
backscattering in a chain even for parameters that do not lead to
backscattering in the two-granule system. 
On the other hand, since pulses are known to remain very
narrow in granular chains~\cite{sen-book,senpreprint,sen-pasi},
the analysis of the dynamics of two granules
provides important insights into the behavior of longer
chains and other larger arrays~\cite{we}.

\section*{Acknowledgments}
This work was supported by the Engineering Research Program of
the Office of Basic Energy Sciences at the U. S. Department of Energy
under Grant No. DE-FG03-86ER13606. 

\appendix
\section{Scaling}
\label{appendixa}
The equations of motion with kinetic friction are in general of the form
\begin{equation}
m\ddot{y}_i(\tau) = \pm \mu_i g  - a[y_i(\tau)-y_j(\tau)]^{n-1}
\end{equation}
where the $y$'s are displacements, $\tau$ is the time, 
$n$ is the exponent in Eq.~(\ref{hertz}), $\mu_i$  is the friction
coefficient, $g$ is the gravitational constant, and $i,j=1,2$. 
If particle
$i$ is frictionless, the first term on the right is absent. 
The initial conditions are $y_i(0)=0$ and $\dot{y}_i(0)=0$ or $v_0$.
We define new variables $x_i$ and $t$ via the relations,
\begin{equation}
y_i=Ax_i, \qquad \tau=Bt,
\end{equation}
in terms of which the equations of motion are
\begin{equation}
\frac{mA}{B^2} \ddot{x}_i(t) = \pm \eta_i  - aA^{n-1}[x_i(t)-x_j(t)]^{n-1}.
\end{equation}
The choices
\begin{equation}
A= \left( \frac{v_0^2m}{a}\right)^{1/n}, \quad 
B= \frac{1}{v_0}\left( \frac{v_0^2m}{a}\right)^{1/n}, \quad 
\eta\equiv \frac{\mu g}{mv_0^2} \left(\frac{v_0^2m}{a}\right)^{1/n}
\end{equation}
lead to the scaled equations of motion used in the text, e.g.
\begin{equation}
\ddot{x}_i(t) = \pm \eta_i - [x_i(t)-x_j(t)]^{n-1},
\end{equation}
with the initial conditions $x_i(0)=0$ and $\dot{x}_i(0)=0$ or $1$.

With hydrodynamic friction the equations of motion are of the form
\begin{equation}
m\ddot{y}_i(\tau) = - \tilde{\gamma}_i \dot{y}_i(\tau) 
- a[y_i(\tau)-y_j(\tau)]^{n-1}
\end{equation}
and, in terms of the new variables, 
\begin{equation}
\frac{A}{B^2} \ddot{x}_i(t) = - \tilde{\gamma}_i\frac{A}{B} \dot{x}_i(t)
- aA^{n-1}[x_i(t)-x_j(t)]^{n-1}.
\end{equation}
Again we choose $A$ and $B$ as above, and set 
\begin{equation}
\gamma_i=\frac{\tilde{\gamma}_i}{mv_0} \left( \frac{v_0^2m}{a}\right)^{1/n}
\end{equation}
to obtain
\begin{equation}
\ddot{x}_i(t) = - \gamma_i \dot{x}_i - [x_i(t)-x_j(t)]^{n-1}
\end{equation}
with initial conditions as above.

\end{document}